\documentclass[twocolumn,showpacs,superscriptaddress,prl]{revtex4}

\usepackage{amsmath, amsthm, amssymb}
\usepackage{graphicx}
\usepackage{dcolumn}
\usepackage{bm}
\usepackage{color} 

\newcommand{\de}[1]{\left( #1 \right)}

\renewcommand{\Re}[1]{{\mathrm{Re}}\de{#1}}

\newcommand{\ket}[1]{| #1 \rangle}
\newcommand{\bra}[1]{\langle #1 }

\begin{document}

\title{Robust-fidelity atom-photon entangling gates in the weak-coupling regime}
\author{Ying Li}
\affiliation{Centre for Quantum Technologies, National University of Singapore, Singapore 117543}
\author{Leandro Aolita}
\affiliation{ICFO-Institut de Ci\`encies Fot\`oniques, Parc Mediterrani
 de la Tecnologia, 08860 Castelldefels (Barcelona), Spain}
\author{Darrick E. Chang}
\affiliation{ICFO-Institut de Ci\`encies Fot\`oniques, Parc Mediterrani
 de la Tecnologia, 08860 Castelldefels (Barcelona), Spain}
\author{Leong Chuan Kwek}
\affiliation{Centre for Quantum Technologies, National University of Singapore, Singapore 117543}
\affiliation{Institute of Advanced Studies (IAS), Nanyang Technological University, Singapore 639673}
\affiliation{National Institute of Education, Nanyang Technological University, Singapore 637616}

\date{\today, version 2}

\begin{abstract}
We describe a simple entangling principle based on the scattering of photons off single emitters in one-dimensional waveguides (or extremely lossy cavities).
The scheme can be applied to polarization- or time-bin- encoded photonic qubits, and features a filtering mechanism that works effectively as a built-in error-correction directive.
This automatically maps imperfections from the dominant sources of errors into heralded losses instead of infidelities, something highly advantageous, for instance, in quantum information applications.
The scheme is thus adequate for high-fidelity maximally entangling gates even in the weak-coupling regime.
These, in turn, can be directly used to store and retrieve photonic-qubit states, thereby completing an atom-photon interface toolbox, or applied to sequential measurement-based quantum computations with atomic memories.
\end{abstract}

\maketitle

\textit{Introduction}.---Photons constitute the most natural system to transport qubits (quantum bits) \cite{Gisi02}.
They have been dubbed ``flying qubits'' for the ease with which they can be sent to distant locations.
On the other hand, due to their stability and long-coherence properties, atoms offer a physical realization of ``stationary qubits.''
Controlled interactions between photons and atoms \cite{Exploring} are thus crucial for quantum networking \cite{QuantNet}.
In this respect, maximally entangling gates stand out.
They are used for state-transfer from atoms to photons \cite{McKeever&Wilk&Keller}, or vice versa \cite{QuantNet} to entangle distant atoms via flying photons \cite{Moehring,QuantNet} or different flying photons via atoms \cite{Wilk&Weber&Lindner&Li} and, ultimately, for measurement-based quantum computations sequentially distributed among hybrid atomic-photonic systems \cite{Anders,Roncaglia}.

The dominant approach to single-atom-single-photon interaction has focused on the {\it strong-coupling regime}, particularly for atoms in high-finesse optical cavities \cite{Exploring,QuantNet,McKeever&Wilk&Keller, Moehring,Wilk&Weber&Lindner&Li}.
There, the coherent interaction between the atom and the cavity mode dominates over cavity leakage and atomic decay.
However, despite remarkable progress \cite{Exploring,QuantNet,McKeever&Wilk&Keller,Moehring,Wilk&Weber&Lindner&Li}, the strong-coupling regime remains challenging for single cavity-emitter setups and poses a formidable obstacle for cascaded arrangements, as required for quantum networks.
An alternative is to exploit the so-called {\it Purcell regime} \cite{Waks&Alexia&Bonato}, where the cavity-atom coupling is stronger than the atomic decay rate, but not the cavity-loss rate.
The cavity is then typically referred to as a {\it bad cavity}, with an enhancement of the atomic spontaneous-emission rate into the cavity output as the main effect (the Purcell effect), instead of coherent oscillations.
This particular form of {\it weak-coupling regime} is less technically demanding and still allows for interesting state manipulations \cite{Waks&Alexia&Bonato}.

In fact, this is exploited in a recent proposal \cite{Chang} where a single quantum emitter is coupled to a one-dimensional (1D) waveguide, which can be thought of as a cavity in the limit of infinite losses, exhibiting tight transverse field confinement.
This confinement induces a strong emitter-field coupling, which, it turns out, can yield very high Purcell factors $P$, indicating the system operates deep in the Purcell regime \cite{Chang}.
With this, an entanglement between flying photons and the emitters can in principle be created via resonant 1D scattering \cite{Kojima,Fan} in the waveguide.
This promising idea has several potential implementations \cite{hollowfiber,nanofiber,Qdots,NVCandWG&NVCandNW}.
Nonetheless, because of emitter decay and finite coupling strengths, all physical setups are restricted to finite $P$.
Moreover, the scattering quality is, in addition, affected by nonzero photonic bandwidths or detunings, and so is the absorption probability, so that the scattering event may not even take place at all.

\begin{figure}[tbp]
\includegraphics[width=8.0 cm]{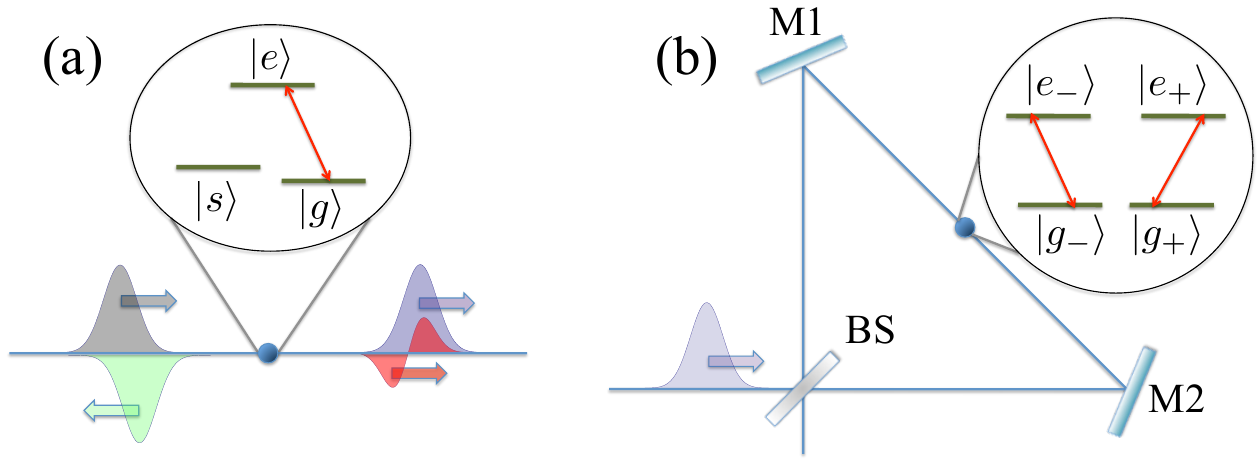}
\caption{
Different 1D scattering setups.
(a) The original arrangement \cite{Chang} uses a three-level emitter, with levels $\vert g\rangle$ and $\vert e\rangle$ coupled via the waveguide, and a third metastable level $\vert s\rangle$ coupled to $\vert e\rangle$ only via classical fields.
In an ideal situation, an incident photon (black) is fully reflected (green), for $\vert g\rangle$, or goes freely through (blue), for decoupled state $\vert s\rangle$.
In a faulty scattering though, there is a transmitted component (red) even for $\vert g\rangle$.
(b) In the present setup the scatterer has twofold degenerate ground and excited states $\vert g_{\pm} \rangle$ and $\vert e_{\pm} \rangle$, respectively, coupled by parallel transitions through orthogonally polarized waveguide photons.
Even for imperfect scattering processes, if the photon is output with the correct polarization, a high-fidelity phase gate is successfully applied on the emitter.
A detection of an incorrectly polarized output, on the other hand, heralds a failure.
A 50/50 beam splitter (BS) and two mirrors (M1 and M2) maximize the probability of success (see text).
}
\label{setup}
\end{figure}

In this paper, we propose a practical scheme for single-emitter-single-photon interfacing that circumvents these limitations.
Physical errors from weak couplings, atomic decay into undesired modes, frequency mismatches, or finite bandwidths of the incident photonic pulses are mapped into heralded photon losses instead of computational errors.
This is a highly desirable feature for quantum communication or computation.
It is achieved with a scattering configuration that swaps the polarization of photons, so that nonscattered photons can automatically be discarded through polarization filtering.
Furthermore, even for faulty processes, e.g., those operating at low $P$, the correct-polarization output photons imprint a phase in the internal state of the emitter.
We exploit this for maximally entangling gates between stationary qubits encoded in the ground states of optical emitters in 1D waveguides and flying qubits encoded in either the polarization or the time of arrival (time bin \cite{timebin}) of photons.
In addition, the gates allow for the storage or retrieval of flying qubits, as well as measurement-based quantum computations sequentially distributed among the single-emitter quantum memories.

\textit{1D photon scattering.}---A two-level emitter, with ground and excited states $\vert g\rangle$ and $\vert e\rangle$, respectively, dipole-coupled to a 1D continuum of electromagnetic modes can, ideally, act as a perfect photon mirror \cite{Fan}.
Specifically, when the excited state is decoupled from any additional channels, incident photons centered in a narrow bandwidth around resonance are fully reflected due to the destructive interference between the reemitted and the (nonabsorbed) incident wave functions.
Technically, for an incident photon in state $\vert \Psi \rangle =\int dz\psi(z,t) \vert z\rangle$, a perfect reflection leads to $\vert \overline{\Psi} \rangle \doteq-\int dz\psi(-z,t) \vert z\rangle$, while a perfectly transmitted (freely propagating) photon remains in $\vert \Psi \rangle$.
Here, $z$ is the spatial coordinate along the waveguide, taken as positive to the right and negative to the left, with the origin $z=0$ at the atom's position; $t$ is the time, with the origin $t=0$ at the scattering instant; $\vert z\rangle$ is the state of a photon at $z$; and $\psi(z,t) \equiv \psi(t-z/c)$ is a normalized wave function, where $c$ is the photonic group velocity inside the waveguide: $c>0$ ($c<0$) for photons propagating to the right (left).
The global minus sign in the definition of $\vert \overline{\Psi} \rangle$ comes from the absorption and subsequent reemission.

Perfect 1D scattering can be used to create emitter-photon qubit entanglement.
Consider an extra metastable level $\vert s\rangle$ decoupled from the waveguide light \cite{Chang} [see Fig. \ref{setup} (a)].
A stationary qubit can then be encoded in the stable atomic manifold, $\{ \vert 0\rangle_{\text{a}} \doteq \vert s\rangle, \vert 1\rangle_{\text{a}} \doteq \vert g\rangle \}$, and a flying qubit in the spatial wave function of single photons, $\{\left\vert 0\right\rangle _{\mathrm{p}}\doteq\left\vert \Psi^{\text{R}}\right\rangle ,\left\vert 1\right\rangle _{\mathrm{p}}\doteq\left\vert \Psi^{\text{L}}\right\rangle\}$, where $\vert \Psi^{\text{R}} \rangle$ and $\vert \Psi^{\text{L}} \rangle$ represent incident wave packets with the same waveform but propagating from left to right and vice versa, respectively.
For the emitter in $\vert 1\rangle_{\text{a}}$, a perfect reflection causes $\vert \Psi^{\text{R(L)}} \rangle \rightarrow \vert \overline{\Psi^{\text{R(L)}}} \rangle\doteq-\vert \Psi^{\text{L(R)}} \rangle$.
Therefore, since $\vert 0\rangle _{\text{a}}$ is decoupled, a perfect process executes the maximally entangling gate $|\mu \rangle_{\text{a}}\vert \varphi\rangle_{\text{p}} \rightarrow (-X_{\text{p}})^{\mu}|\mu \rangle_{\text{a}}\vert \varphi\rangle_{\text{p}}$, where $\ket{\varphi}_{\text{p}}$ is any photonic-qubit state, $X_{\text{p}}$ the corresponding first Pauli matrix, and $\mu =0$ or $1$.

In practice, however, the reemitted amplitude is weaker than the incident one and cannot cancel it.
There is always a transmitted part \cite{Chang}.
For incident state $\vert \Psi \rangle$, the photon comes out in $\vert \Phi \rangle =\vert \Phi_{\text{t}} \rangle +\vert \Phi_{\text{r}} \rangle$, with transmitted and reflected components $\vert \Phi _{\text{t}} \rangle =\int dz\phi_{\text{t}}(z,t) \vert z\rangle $ and $\vert \Phi_{\text{r}} \rangle =\int dz\phi_{\text{r}}(-z,t) \vert z\rangle$, respectively, with \cite{Kojima,Fan,Chang}
\begin{subequations}
\label{wavefuncs}
\begin{align}
\label{wavefunct}
\phi _{\mathrm{t}}(z,t) &=\psi(z,t)+\phi _{\mathrm{r}}(z,t),\\
\label{wavefuncr}
\phi_{\mathrm{r}}(z,t) &=-\frac{\Gamma _{\text{1D}}}{2}\int_{0}^{t-z/c}dt^{\prime } \nonumber \\
&\times e^{-i(\omega _{0}-i\Gamma /2)(t-z/c-t^{\prime })}\psi(0,t^{\prime }).
\end{align}
\end{subequations}
Here, $\Gamma\doteq\Gamma _{\text{1D}}+\Gamma^{\prime }$ is the total atomic decay rate, with $\Gamma _{\text{1D}}$ ($\Gamma^{\prime}$) the rate of atomic decay into the waveguide (out of the waveguide, e.g., emission into free space, or nonradiative dissipation), and $\omega_0$ is the atomic transition frequency.
$\ket{\Phi}$ refers to the state-component left in the waveguide, so it is normalized only when the Purcell factor $P\doteq\Gamma_{\text{1D}}/\Gamma'$ is infinite.
In particular, for finite $\Gamma^{\prime}$ and $\Gamma_{\text{1D}}\to\infty$, a Dirac delta appears in the integrand of Eq. \eqref{wavefuncr}, so that $\phi_{\text{r}}(z,t) =-\psi(0,t-z/c) \equiv -\psi(t-z/c) \equiv -\psi(z,t)$ and one has a perfect reflection: $\ket{\Phi}=\ket{\Phi_{\text{r}}} =\ket{\overline{\Psi}}$.
Accordingly, the probability of photon loss is $\kappa \doteq 1-\bra{\Phi}\ket{\Phi}$.

Apart from $P$, another relevant figure of merit is the reflection fidelity $f\doteq -\bra{\overline{\Psi}} \ket{\Phi_{\text{r}}}$, which measures how close to a perfect reflection the process is and can also be affected by frequency detunings or nonzero photonic bandwidths.
In terms of $P$ and $f$, the probability of photon transmission and reflection are given, respectively, by \cite{Kojima,Fan,Chang} the transmittance $\mathcal{T} \doteq \bra{\Phi_{\text{t}}} \ket{\Phi_{\text{t}}} =1-\Re{f}[2-1/(1+P^{-1})]$ and the reflectance $\mathcal{R} \doteq \bra{\Phi_{\text{r}}} \ket{\Phi_{\text{r}}} =\Re{f}/(1+P^{-1})$.
A maximally entangling gate can only be obtained for $P\to\infty$ and $f=1$, because only then does one have $\mathcal{R}=1$ (so that $\vert \Psi^{\text{R(L)}} \rangle \rightarrow -\vert \Psi^{\text{L(R)}} \rangle$).
The lower $\mathcal{R}$ is, the lower the fidelity of the resulting gate.

\par As a simple example, imagine an incident photon spontaneously emitted, at rate $\gamma$, by a distant emitter.
In this case, the photon has a half-exponential waveform of bandwidth $\gamma$.
Equation \eqref{wavefuncr} is then immediately integrated to yield $f=(1 +P^{-1} +\gamma /\Gamma_{\text{1D}} -i2\delta /\Gamma_{\text{1D}} )^{-1}$, where $\delta$ is the detuning from $\omega_0$.
Notice that even if $P\to\infty$ and $\delta\approx0$, already for $\gamma \approx \Gamma_{\text{1D}}$, $f$ (and therefore also $\mathcal{R}$) decreases to $1/2$.
This would indeed be the case when emitter and scatterer are of the same species.
More generally, this limitation is a serious drawback for short pulses, as those used in time-bin qubits \cite{Gisi02,timebin}.

\textit{High-fidelity interaction from imperfect processes.}---Consider now a four-level emitter, with degenerate ground and excited states $\vert g_{\pm} \rangle$ and $\vert e_{\pm} \rangle $ [see Fig. \ref{setup} (b)].
These are coupled via parallel dipole transitions $\vert g_{\pm} \rangle \leftrightarrow \vert e_{\pm} \rangle$, associated with the absorption from, or emission to, the waveguide of $\sigma^{\pm}$-polarized photons.
$\sigma^{+}$ and $\sigma^{-}$ denote two orthogonal polarizations as, for instance, the right- and left-handed circular polarizations along the waveguide.
The waveguide is taken as the atomic quantization axis.
An incident photon of spatial wave function $\ket{\Psi}$ and polarization $\sigma^{\pm}$ scatters as
\begin{subequations}
\label{transf}
\begin{align}
\ket{g_{\pm}}\ket{\Psi}\ket{\sigma^\pm}
\rightarrow \ket{g_{\pm}}\ket{\Phi}\ket{\sigma^\pm},\\
\ket{g_{\mp}}\ket{\Psi}\ket{\sigma^\pm}
\rightarrow \ket{g_{\mp}}\ket{\Psi}\ket{\sigma^\pm}.
\end{align}
\end{subequations}

If, instead, the photon is in the linear-polarization state $\ket{h} \doteq (\ket{\sigma^{+}}+\ket{\sigma^{-}})/\sqrt{2}$, transformations \eqref{transf} yield
\begin{eqnarray}
\label{transfH}
\nonumber
&\ket{g_{\pm}}\ket{\Psi}\ket{h} \rightarrow\\
&\dfrac{1}{2}\ket{g_{\pm}} \big[(\ket{\Phi}+\ket{\Psi})\ket{h}
\pm(\ket{\Phi}-\ket{\Psi})\ket{v}\big],
\end{eqnarray}
where $\ket{v} \doteq (\ket{\sigma^{+}}-\ket{\sigma^{-}})/\sqrt{2}$ is the vertical linear-polarization state.
Now, the scattering generates a $v$-polarized component.
More importantly, while for $h$-polarized outgoing photons nothing happens to the emitter, a state-dependent $\pi$-phase shift on the emitter accompanies the $v$-polarized component of Eq. \eqref{transfH}.
This phase shift will be the basis of our entangling gates.

To maximize the $v$-polarized component, the input photon is coherently split into two halves that scatter simultaneously, each incident from a different side [see Fig. \ref{setup} (b)].
Next, the reflected and transmitted components of each half are coherently joined back into a single packet, which exits the beam splitter (BS) through the same mode it was input.
Then, $(\ket{\Psi}-\ket{\Phi})/2=-\vert \Phi_{\text{r}} \rangle$, and discarding the $h$-polarized output from Eq. \eqref{transfH}, one gets
\begin{eqnarray}
\label{transreduced}
\ket{\varphi}_{\text{a}}\ket{\Psi}\ket{h} \rightarrow
-Z_{\text{a}}\ket{\varphi}_{\text{a}}\vert \Phi_{\text{r}} \rangle \ket{v},
\end{eqnarray}
where $\ket{\varphi}_{\text{a}}$ is any atomic-qubit state in the basis $\{\left\vert 0\right\rangle _{\text{a}}\doteq\left\vert g_-\right\rangle,\left\vert 1\right\rangle _{\text{a}}\doteq\left\vert g_+\right\rangle\}$.
For perfect scattering processes, $\ket{\Phi_{\text{r}}}=-\ket{\Psi}$, and therefore the success probability $p_{\text{s}}\doteq\bra{\Phi_{\text{r}}}\ket{\Phi_{\text{r}}}$ is $1$.
No photon is lost then.
On the other hand, for imperfect processes, with $P<\infty$, $\ket{\Phi_{\text{r}}}\neq-\ket{\Psi}$, and output photons with $h$ polarization are detected.
These are discarded, and the corresponding gate runs fail.
However, the important thing is that the fidelity of gate \eqref{transreduced} is not altered; only $p_{\text{s}}$ is.
Next, we show how to exploit the successful $Z_{\text{a}}$ gates for high-fidelity entangling schemes.

\textit{Entangling gate for time-bin flying qubits.}---The first photonic-qubit encoding we consider is the time of arrival \cite{timebin}, consisting of incident pulses that arrive either at some ``early'' scattering time $t_e$, defined as the state $\vert \Psi_{t_e} \rangle$, at some ``later'' time $t_l>t_e$ \cite{timebin}, defined as $\vert \Psi_{t_l} \rangle$, or in any superposition of the latter two states.
The qubit basis is $\{ \vert 0\rangle_{\text{p}} \doteq \vert \Psi_{t_e} \rangle ,\vert 1\rangle_{\text{p}} \doteq \vert \Psi_{t_l} \rangle \}$.
The final ingredient of the protocol is the application of a Hadamard gate $H_{\text{a}}$ to the atomic qubit between $t_e$ and $t_l$: if the photon arrives at $t_e$, the emitter undergoes first $Z_{\text{a}}$ and then $H_{\text{a}}$, whereas if it arrives at $t_l$, the order of the gates is reversed.
Since $Z_{\text{a}}$ and $H_{\text{a}}$ do not commute, the overall stationary-qubit gate is controlled by the flying qubit's state.
The composite unitary transformation
\begin{equation}
\label{compositegate}
U_{\text{ap}}=\left\vert 0\right\rangle _{\text{p}}\left\langle 0\right\vert \otimes H_{\text{a}}Z_{\text{a}}
+\left\vert 1\right\rangle _{\text{p}}\left\langle 1\right\vert\otimes Z_{\text{a}}H_{\text{a}}
\end{equation}
 is local-unitarily equivalent to the well-known controlled-phase gate, and is therefore also maximally entangling.

\begin{figure}[t]
\includegraphics[width=7.0 cm]{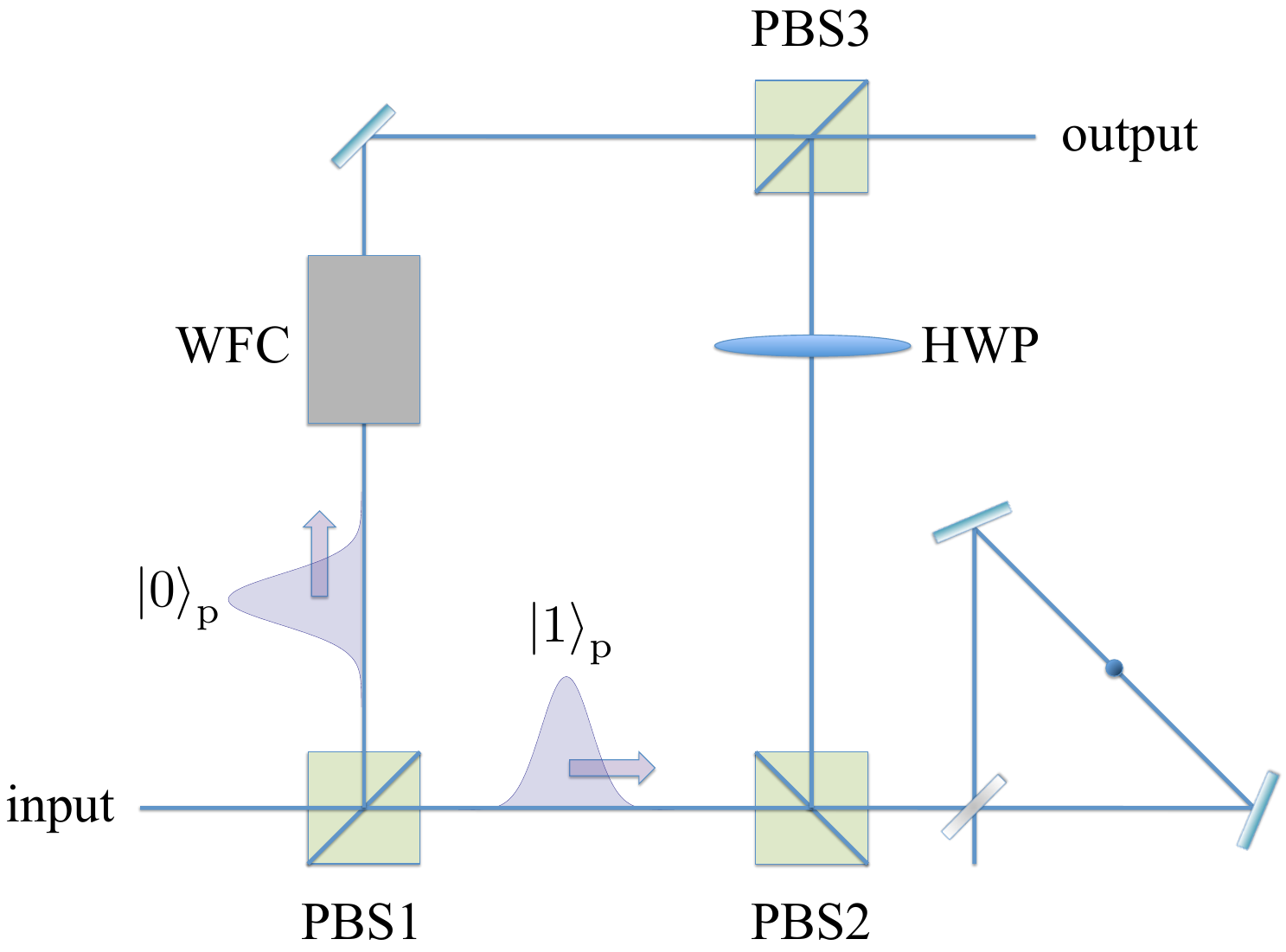}
\caption{
Interferometric setup for polarization flying qubits.
Only component $\ket{1}_{\text{p}}$ interacts with the scatterer.
For successful interactions, $\ket{1}_{\text{p}}$ is reflected up the interferometer by PBS2 and joins $\ket{0}_{\text{p}}$ at PBS3.
At the output, the maximally entangling gate \eqref{compositegatepol} between the flying and the stationary qubits is implemented.
For unsuccessful interactions, $\ket{1}_{\text{p}}$ comes back out through PBS2 and PBS1, which heralds a gate failure.
Legends: PBS = polarizing BS; HWP = half-wave plate; WFC = waveform corrector.
See text.
}
\label{polarization_setup}
\end{figure}

\textit{Entangling gate for polarization flying qubits.}---The other encoding considered is photon polarization.
We define it by $\{\left\vert 0\right\rangle _{\text{p}}\doteq\left\vert v\right\rangle ,\left\vert 1\right\rangle _{\text{p}}\doteq\left\vert h\right\rangle\}$.
In this case, the active application of $H_{\text{a}}$ is replaced with the passive interferometer shown in Fig. \ref{polarization_setup}.
First, the $\left\vert 0\right\rangle _{\text{p}}$ and $\left\vert 1\right\rangle _{\text{p}}$ components of an incident photon are spatially split by a polarizing beam splitter (PBS).
$\left\vert 1\right\rangle _{\text{p}}$ goes through both PBS1 and PBS2 towards the scattering setup, whereas component $\left\vert 0\right\rangle _{\text{p}}$ is reflected up the other arm of the interferometer by PBS1.
For unsuccessful events, $\left\vert 1\right\rangle _{\text{p}}$ exits the scattering setup with the same polarization, $h$.
It is therefore transmitted back through PBS2 and PBS1, and detected in $h$ as before, heralding the failure of the gate run.
On the other hand, for successful $Z_{\text{a}}$ gates the polarization is swapped.
Since it is then $v$-polarized, the pulse is reflected up by PBS2, after which it is rotated back to $\ket{h}$ by a half-wave plate (HWP).
Finally, $\left\vert 0\right\rangle _{\text{p}}$ and $\left\vert 1\right\rangle _{\text{p}}$ are rejoined by PBS3.
At the output of PBS3, the total composite unitary transformation is the controlled-phase gate
\begin{equation}
\label{compositegatepol}
U_{\text{ap}}=\left\vert 0\right\rangle _{\text{p}}\left\langle 0\right\vert \otimes \openone_{\text{a}}
+\left\vert 1\right\rangle _{\text{p}}\left\langle 1\right\vert\otimes Z_{\text{a}}.
\end{equation}

For successful events of imperfect processes, i.e., where the polarization is swapped but $\ket{\Phi_{\text{r}}}\neq-\ket{\Psi}$ in Eq. \eqref{transreduced}, the spatial wave functions of the packets meeting at PBS3 no longer coincide.
Therefore, unless the waveforms are matched, the output polarization qubit may be correlated with different spatial states.
An experimentally relevant situation where this can be easily overcome is for incident photons with bandwidth much narrower than $\Gamma_{1D}$, so that $\psi(z,t)$ is approximated in Eq. \eqref{wavefuncr} by a plane wave.
In this case the waveform associated with $\left\vert 1\right\rangle _{\text{p}}$ is $\ket{\Phi_{\text{r}}}\approx-k\ket{\Psi}$, with $|k|<1$.
To compensate for this, a waveform corrector (WFC) in the $\left\vert 0\right\rangle _{\text{p}}$ arm maps $\ket{\Psi}$ to $k\ket{\Psi}$.
This slightly decreases the overall success probability, but leaves the fidelity intact.
When the photon-atom detuning $\delta$ is zero, $k\in[0,1)$, and the WFC consists of an attenuator (e.g., a BS) of transmissivity $k$.
If $\delta\neq0$, $k\in\mathbb{C}$, and the WFC simply includes also a phase modulator \cite{comment0}.
In the general situation $\ket{\Phi_{\text{r}}}\neq\ket{\Psi}$, the WFC can be realized by a second scattering block, identical to that of Fig. \ref{setup} (b), but with the emitter permanently in $\ket{g_+}$ (or $\ket{g_-}$), preceded by a quarter wave plate to rotate $\ket{0}_{\text{p}}$ to $\ket{\sigma^+}$ (or $\ket{\sigma^-}$).
With this, the associated wave packet is mapped from $\ket{\Psi}$ to $\ket{\Phi_{\text{r}}}$ without entangling with the second scatterer, thus achieving the desired matching.

\textit{Quantum memories and quantum computations.}---These maximally entangling gates, together with single-qubit gates and measurements, allow for efficient measurement-based quantum computations sequentially distributed (by the flying qubits) among distant nodes of a quantum network \cite{Anders,Roncaglia}.
The underlying model is the one-way quantum computer \cite{Oneway}, but the approaches of Refs. \cite{Anders,Roncaglia} have the advantages that
(i) only the relevant pieces of the cluster are created (and almost immediately consumed) \cite{Roncaglia},
(ii) the total number of required stationary qubits is drastically smaller than in the one-way model \cite{Roncaglia},
and (iii) every flying qubit interacts with at most two stationary ones, and typically with only one \cite{Anders,Roncaglia}.

Since these models are universal \cite{Anders,Roncaglia}, they include the creation of multipartite entanglement among different scatterers, or simply the storage, and later retrieval, of flying qubits, so that each emitter works as a quantum memory.
The storage consists essentially of maximally entangling the incident photon with the emitter, with a subsequent measurement on the outgoing photon.
The retrieval, in turn, is done by maximally entangling a second photon with the emitter qubit, in the stored state, followed by a measurement on the emitter.
As a result, the second photon takes the stored state away with it.

\textit{Feasibility.}---For artificial solid-state emitters, such as quantum dots \cite{Qdots} or nitrogen-vacancy centers \cite{NVCandWG&NVCandNW}, coupled to photonic nanowires or photonic-crystal waveguides, $P>20$ has been demonstrated \cite{Qdots}.
This corresponds to $p_{\text{s}}>0.95$, thus providing a candidate for implementation.
In addition, the observed decay rate is $\Gamma_{\text{1D}}>1\text{ GHz}$ \cite{Qdots}, so that photons with pulse durations of around tens of nanoseconds can scatter with excellent reflection fidelities.
For the time-bin scheme, the Hadamard gate needed between the $\ket{0}_{\text{p}}$ and $\ket{1}_{\text{p}}$ components can be implemented in picoseconds \cite{Picos}.
Then, the total duration of the gate would be comparable to the coherence time of bare quantum dots \cite{dephasing,Picos}.
Nevertheless, by quantum-controlling the surrounding nuclear-spin bath, this can be enhanced by up to two orders of magnitude \cite{Reilly&Lange}.

Other potential setups are atoms coupled to hollow fiber cores \cite{hollowfiber} or ultrathin nanofibers \cite{nanofiber}.
The modest Purcell factors ($P\lesssim 1$) there are enough to yield $p_{\text{s}}\lesssim 0.5$.
In addition, whereas coupling atoms to fibers is still challenging, the techniques progress remarkably fast \cite{hollowfiber, nanofiber}, and atoms provide coherence times as high as seconds.
Finally, several groups have demonstrated the strong coupling of a single atom without a cavity with a tightly focused laser \cite{FreeSpace}.
When this is properly mode-matched to the atomic emission, the description also resembles 1D atom-photon scattering.

\textit{Heralded losses versus infidelities.}---Turning errors into detectable losses is advantageous for quantum information, as low efficiencies are typically simpler to handle than low fidelities.
For example, the most optimistic thresholds of error rate per gate for fault-tolerant quantum computing are below $3\%$ \cite{FTQC}.
In contrast, the one-way quantum computer \cite{Oneway}, as well as its sequential counterparts \cite{Anders, Roncaglia} considered here, can cope with loss rates close to $50\%$ \cite{Loss}, and heralded gate-failure rates above $90\%$ \cite{DFTQC}.
Another example is long-distance quantum communication with quantum repeaters \cite{Repeat}.
There, if an entangled pair is lost, one reestablishes the repeater link simply by distributing a new pair, but if the distributed pairs are faulty, their infidelity propagates exponentially with the number of links.

\textit{Conclusion}---We have proposed a simple scattering configuration for photons and optical emitters in 1D waveguides.
This allows for probabilistic maximally entangling gates between stationary and flying qubits.
Faulty interactions are tagged with an orthogonal output polarization, which can be immediately discarded, rendering a built-in error-heralding mechanism.
This turns gate infidelities into heralded losses.
The gates then either succeed with perfect fidelity or fail in a heralded manner, but are in principle never faulty.
We have estimated success probabilities for current setups that range from $\lesssim 50\%$ to as high as $95\%$.
The gates are thus adequate for the storage or retrieval of flying-qubit states, and for measurement-based multiparty-state preparations, or quantum computations, sequentially distributed among distant emitters.

\begin{acknowledgements}
We acknowledge support from the National Research Foundation and Ministry of Education, Singapore. L.A. acknowledges support from Fundaci\'on Juan de la Cierva, and D.E.C. from Fundaci\'o Privada Cellex Barcelona.
\end{acknowledgements}


\begin{thebibliography}{20}
\bibitem{Gisi02} N. Gisin \textit{et al.}, 
Rev. Mod. Phys. \textbf{74}, 1455 (2002).

\bibitem{Exploring} S. Haroche and J. M. Raimond, \textit{Exploring the Quantum: Atoms, Cavities and Photons} (Oxford University Press, Oxford, United Kingdom, 2006); K. Hammerer
A. S. S\o rensen and E. S. Polzik, Rev. Mod. Phys. \textbf{82}, 1041 (2010).

\bibitem{QuantNet} J. I. Cirac \textit{et al.}, 
Phys. Rev. Lett. {\bf 78}, 3221 (1997); H. J. Kimble, Nature \textbf{453}, 1023 (2008); S. Ritter \textit{et al.}, Nature \textbf{484}, 195 (2012).

\bibitem{McKeever&Wilk&Keller} J. McKeever \textit{et al.}, Nature \textbf{425}, 268 (2004); 
M. Keller \textit{et al.}, Nature \textbf{431}, 1075 (2004); 
T. Wilk \textit{et al.}, 
Phys. Rev. Lett. {\bf 98}, 063601 (2007).


\bibitem{Moehring} D. L. Moehring \textit{et al.}, 
Nature \textbf{449}, 68 (2007).

\bibitem{Wilk&Weber&Lindner&Li} T. Wilk \textit{et al.}, 
Science \textbf{488}, 317 (2007);
B. Weber {\it et al.}, 
Phys. Rev. Lett. {\bf 102} 030501 (2009);
N. H. Lindner and T. Rudolph, Phys. Rev. Lett. {\bf 103}, 113602 (2009); 
Y. Li, L. Aolita, and L. C. Kwek, Phys. Rev. A {\bf 83}, 032313 (2011).

\bibitem{Anders}
J. Anders {\it et al.}, 
Phys. Rev. A {\bf 82}, 020301(R) (2010);
E. Kashefi {\it et al.}, 
Electron. Notes Theor. Comput. Sci. {\bf 249}, 307 (2009).

\bibitem{Roncaglia}
A. J. Roncaglia {\it et al.}, Phys. Rev. A {\bf 83}, 062332 (2011).

\bibitem{Waks&Alexia&Bonato}
See E. Waks and J. Vuckovic, Phys. Rev. Lett. {\bf 96}, 153601 (2006);
A. Auff\`{e}ves-Garnier {\it et al.}, 
Phys. Rev. A {\bf 75}, 053823 (2007);
C. Bonato \textit{et al.}, 
Phys. Rev. Lett. \textbf{104}, 160503 (2010); and references therein.


\bibitem{Chang} D. E. Chang \textit{et al.}, 
Nature Phys. \textbf{3}, 807 (2007).

\bibitem{Kojima} K. Kojima \textit{et al.}, 
Phys. Rev. A \textbf{68}, 013803 (2003).

\bibitem{Fan} J. T. Shen, and S. Fan, Opt. Lett. \textbf{30}, 2001 (2005).


\bibitem{hollowfiber} C. A. Christensen {\it et al.}, 
Phys. Rev. A \textbf{78}, 033429 (2008)
; M. Bajcsy {\it et al.}, 
Phys. Rev. Lett. \textbf{102}, 203902 (2009)
.

\bibitem{nanofiber} G. Sagu\'{e} {\it et al.}, 
Phys. Rev. Lett. \textbf{99}, 163602 (2007);
E. Vetsch {\it et al.}, 
Phys. Rev. Lett. \textbf{104}, 203603 (2010).

\bibitem{Qdots} T. Lund-Hansen, {\it et al.},
Phys. Rev. Lett. \textbf{101}, 113903 (2008); J. Gao, F. W. Sun, and C. W. Wong, Appl. Phys. Lett. \textbf{93}, 151108 (2008); J. Claudon {\it et al.}, 
Nature Photon. \textbf{4}, 174 (2010).


\bibitem{NVCandWG&NVCandNW}
T. M. Babinec {\it et al.}, 
Nature Nanotech. \textbf{5}, 195 (2010)

\bibitem{timebin} J. Brendel {\it et al.}, 
Phys. Rev. Lett. \textbf{82}, 2594 (1999).

\bibitem{comment0}
In addition, the WFC may also include a delay to make $\left\vert 0\right\rangle _{\text{p}}$ arrive simultaneously with $\left\vert 1\right\rangle _{\text{p}}$ at PBS3.

\bibitem{dephasing} J. R. Petta {\it et al.}, Science \textbf{309}, 2180 (2005).

\bibitem{Picos} J. Berezovsky {\it et al.}, Science \textbf{320}, 349 (2008).

\bibitem{Reilly&Lange} D. J. Reilly {\it et al.}, 
Science {\bf 321}, 817 (2008); G. de Lange {\it et al.}, 
Sci. Rep. {\bf 2:382}, DOI:10.1038/srep00382 (2012).

\bibitem{FreeSpace} B. Darqui\'e {\it et al.}, Science \textbf{309}, 454 (2005); M. K. Tey 
{\it et al.}, Nature Phys. \textbf{4}, 924 (2008); G. Wrigge {\it et al.}, Nature Phys. \textbf{4}, 60 (2008); G. H\'etet {\it et al.}, Phys. Rev. Lett. \textbf{107}, 133002 (2011).

\bibitem{FTQC} E. Knill, Nature \textbf{434}, 39 (2005).

\bibitem{Oneway} R. Raussendorf and H. J. Briegel, Phys. Rev. Lett. \textbf{86}, 5188 (2001).

\bibitem{Loss} M. Varnava, D. E. Browne, and T. Rudolph, Phys. Rev. Lett. \textbf{97}, 120501 (2006).

\bibitem{DFTQC} Y. Li {\it et al.}, 
Phys. Rev. Lett. \textbf{105}, 250502 (2010);
K. Fujii and Y. Tokunaga, Phys. Rev. Lett. \textbf{105}, 250503 (2010).

\bibitem{Repeat}
H.-J. Briegel {\it et al.}, Phys. Rev. Lett. \textbf{81}, 5932 (1998).

\end{thebibliography}
\end{document}